\documentclass{iaus}
\usepackage{graphicx}

\title[X-Rays in T Tauri Stars] 
{Accretion and Outflow-Related X-Rays \break in T Tauri Stars}

\author[M. G\"udel \etal]   
{Manuel G\"udel$^{1,2}$,  
Kevin Briggs$^1$, 
Kaspar Arzner$^1$,  
Marc Audard$^{3}$,
J\'er\^ome Bouvier$^4$,
Catherine Dougados$^4$,
Eric Feigelson$^5$,
Elena Franciosini$^6$,
Adrian Glauser$^1$,
Nicolas Grosso$^4$
  \thanks{Present address: Observatoire Astronomique de Strasbourg, 11 rue de l'universit\'e, 67000 Strasbourg, France}, 
Sylvain Guieu$^4$
  \thanks{Present address: Spitzer Science Center, California Institute of Technology, Mail Code 220-6, Pasadena, CA 91125, USA},
Fran\c{c}ois M\'enard$^4$,
Giusi Micela$^6$,
Jean-Louis Monin$^4$,
Thierry Montmerle$^4$,
Deborah  Padgett$^7$,
Francesco Palla$^8$,
Ignazio Pillitteri$^{6,9}$,
Thomas Preibisch$^{10}$,
Luisa Rebull$^7$,
Luigi Scelsi$^{6,9}$,
Bruno Silva$^{11}$,
Stephen Skinner$^{12}$,
Beate Stelzer$^6$
\and 
Alessandra Telleschi$^{1}$
}
 
\affiliation{
$^1$     Paul Scherrer Institut, W\"urenlingen and Villigen, 5232 Villigen PSI, Switzerland
              \break email: guedel@astro.phys.ethz.ch
	      \\[\affilskip]
$^2$     Max-Planck-Institute for Astronomy, K\"onigstuhl 17, 69117 Heidelberg, Germany
	      \\[\affilskip]
$^3$     Integral Science Data Centre, Ch. d'Ecogia 16, 1290 Versoix, and
         Geneva Observatory, University of Geneva, Ch. des Maillettes 51, 1290 Sauverny, Switzerland
              \\[\affilskip]
$^4$     Laboratoire d'Astrophysique de Grenoble, Universit\'e Joseph Fourier - CNRS,  BP 53, 38041 Grenoble Cedex, France 
	      \\[\affilskip]
$^5$     Department of Astronomy \& Astrophysics, Penn State University, 525 Davey Lab, University Park, PA 16802, USA 
	      \\[\affilskip]
$^6$     INAF - Osservatorio Astronomico di Palermo, Piazza del Parlamento 1, 90134 Palermo, Italy
	      \\[\affilskip]
$^7$     Spitzer Science Center, California Institute of Technology, Mail Code 220-6, Pasadena, CA 91125, USA 
	      \\[\affilskip]
$^8$     INAF - Osservatorio Astrofisico di Arcetri, Largo Enrico Fermi, 5, 50125 Firenze, Italy  
	      \\[\affilskip]
$^9$     Dipartimento di Scienze Fisiche ed Astronomiche, Universit\`a di Palermo, Piazza del Parlamento 1, 90134 Palermo, Italy  
              \\[\affilskip]
$^{10}$  Max-Planck-Institut f\"ur Radioastronomie, Auf dem H\"ugel 69,  53121 Bonn, Germany
	      \\[\affilskip]
$^{11}$  Centro de Astrof\'{\i}sica da Universidade do Porto, Rua das Estrelas, 4150 Porto, and
         Departamento de Matem\'atica Aplicada, Faculdade de Ci\^ecias da Universidade do Porto, 4169 Porto, Portugal
	      \\[\affilskip] 	    
$^{12}$  CASA, UCB 389, University of Colorado, Boulder, CO 80309-0389, USA 
}

\pubyear{2007}
\volume{243}  
\pagerange{1--8}
\date{?? and in revised form ??}
\jname{Star-Disk Interactions in Young Stars}
\editors{J. Bouvier \& I. Appenzeller, ed.}
\begin{document}

\maketitle

\begin{abstract}
We report on accretion- and outflow-related X-rays
from T Tauri stars, based on results from the "XMM-Newton Extended Survey 
of the Taurus Molecular Cloud." X-rays potentially form in shocks
of accretion streams near the stellar surface, although we 
hypothesize that direct interactions between the streams and magnetic 
coronae may occur as well. We report on the discovery of a "soft excess" in 
accreting  T Tauri stars supporting these scenarios. We further discuss a new 
type of X-ray source in jet-driving T Tauri stars. It shows a strongly 
absorbed coronal component and a very soft, weakly absorbed component 
probably related to shocks in microjets. The excessive coronal absorption 
points to dust-depletion in the accretion streams. 

\end{abstract}

\firstsection 
\section{Introduction}

Classical and weak-lined T Tauri stars (CTTS and WTTS)
are pre-main sequence stars showing vigorous X-ray
emission with X-ray luminosities ($L_{\rm X}$) near the empirical
saturation limit found for main-sequence (MS) stars,
$L_{\rm X}/L_{\rm bol} \approx 10^{-3.5}$. 
Consequently, X-ray emission from T Tauri stars has been attributed 
to solar-like coronal activity. However,
both accretion and outflows/jets may  contribute
to or alter  X-ray production in the magnetic
environment of strongly accreting
young stars. X-rays thus provide an important
diagnostic for the detection and study of accretion
and outflow activity.

Support for X-ray {\it suppression} in CTTS has been reported from X-ray photometry.
CTTS are, on average, less X-ray luminous than WTTS (e.g., \cite[Strom \& Strom 1994]{strom94}, 
\cite[Neuh\"auser \etal\ 1995]{neuhauser95}). This finding has  been partly supported by
recent, deep surveys of the Orion Nebula Cluster (\cite[Getman \etal\ 2005]{getman05}, 
\cite[Preibisch \etal\ 2005]{preibisch05}).

On the other hand, X-rays may be {\it generated} by surface accretion shocks  
(\cite[Lamzin 1999]{lamzin99}). If gas collides with the stellar photosphere in free fall,
shocks heat it to a few million K. Given the appreciable accretion rates,
high shock densities of order $10^{12}- 10^{14}$~cm$^{-3}$ are expected, as first
reported for the CTTS TW Hya (\cite[Kastner \etal\ 2002]{kastner02}). X-rays may
similarly be produced in shocks forming in jets (\cite[Raga \etal\ 2002]{raga02}).

The {\it XMM-Newton Extended Survey of the Taurus Molecular Cloud (XEST)}  
(\cite[G\"udel \etal\ 2007a]{guedel07a}) has provided new insights
into these issues. XEST covers the most populated  $\approx$5~sq. deg of the
Taurus star-forming region. The average on-axis detection 
limit is $\approx 10^{28}$~erg~s$^{-1}$ for lightly absorbed objects, sufficient
to detect about half of the observed brown dwarfs (\cite[Grosso \etal\ 2007a]{grosso07a}).
We discuss X-ray results relevant for  the star-disk interface in
which accretion occurs and where (parts of) the jets may be accelerated.

\section{Accretion and the ``X-ray Soft Excess''}

\subsection{Accretors in XEST}\label{accretors}

\begin{figure}
 \vskip -0.1truecm
 \hbox
 {
 \includegraphics[angle=-0,width=7.3cm]{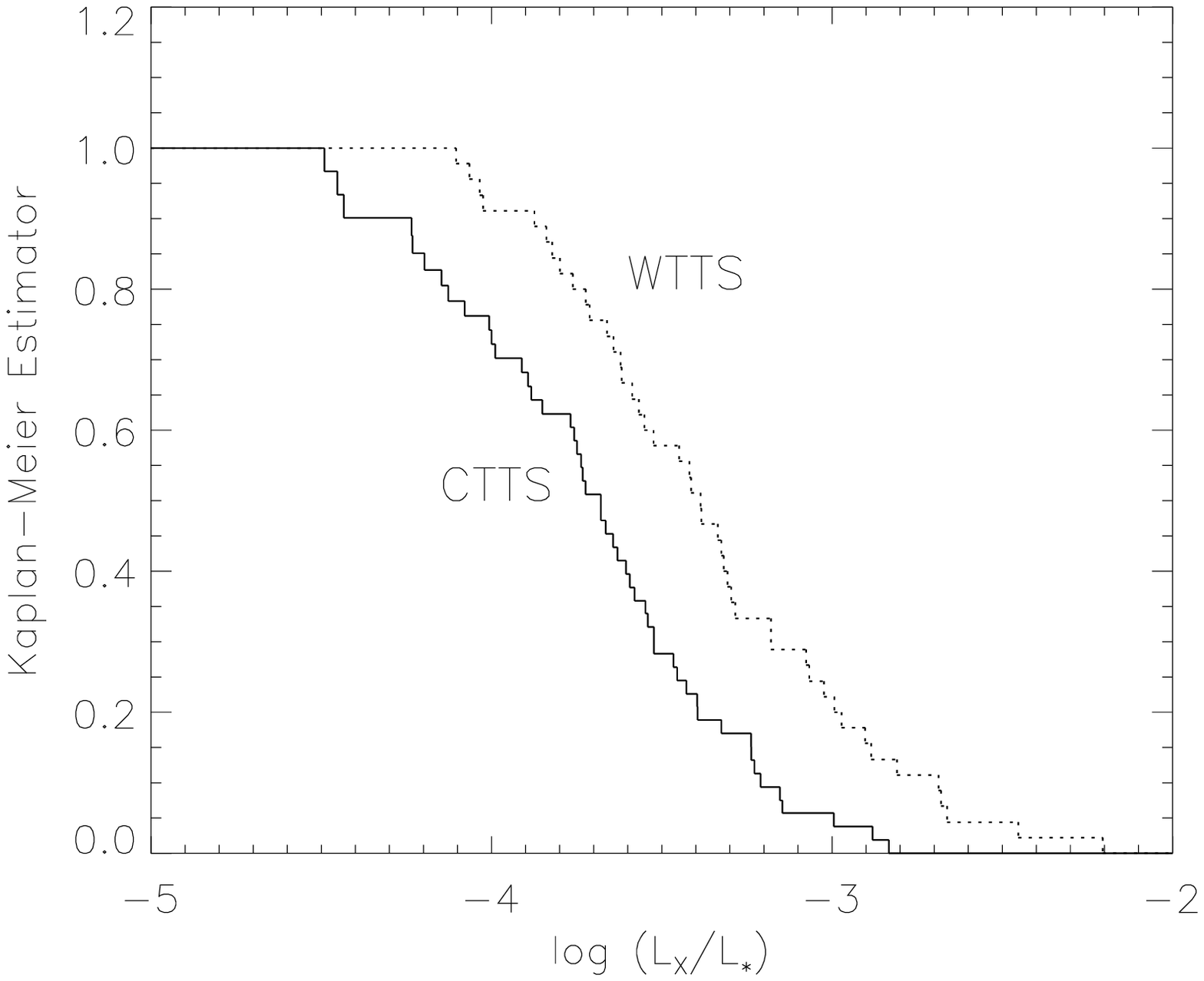}
 \includegraphics[angle=-90,width=5.85cm]{guedelfig1b.ps}
 }
 \vskip -5.2truecm
 \caption{{\it Left (a)}: The cumulative distribution of the $L_{\rm X}/L_{\rm bol}$ ratio for CTTS (solid) and WTTS (dotted)
           are different at the $>99.97$\% level (from \cite[Telleschi \etal\ 2007a]{telleschi07a}).   
	    -- {\it Right (b)}: The O\,{\sc vii} triplet of T Tau, showing a strong forbidden line at 22.1~\AA\ (after
	    \cite[G\"udel \etal\ 2007c]{guedel07c}).}\label{fig:accretion}
\end{figure}

\cite{telleschi07a} present statistical X-ray studies of the  XEST CTTS and WTTS samples.
The WTTS sample is essentially complete (all but one of the 50 surveyed objects detected),
and the CTTS sample is nearly complete (85\% of the 65 surveyed objects detected). The 
luminosity deficiency of CTTS is confirmed. More precisely, CTTS are statistically less luminous 
by a factor of $\approx 2$ both in $L_{\rm X}$ and in $L_{\rm X}/L_{\rm bol}$ (Fig.~\ref{fig:accretion}a) while the 
$L_{\rm bol}$ distributions of the two samples are drawn from the same parent population.

On the other hand, \cite{telleschi07a} report the average electron temperature, $T$, in the
X-ray sources of CTTS  to be higher than in WTTS, irrespective of the gas absorption column 
density ($N_{\rm H}$). For WTTS, a trend also seen in MS
stars is recovered, in the sense that  the electron temperature increases with  $L_{\rm X}$.
Such trends are expected in stochastic-flare heated coronae (\cite[Telleschi \etal\ 2005]{\telleschi05}),
but a similar trend is absent in CTTS in which the average temperature remains high for 
all activity levels.

In contrast to the apparent X-ray deficiency, a trend toward an ultraviolet excess is found in
CTTS based on the Optical Monitor (OM) data (\cite[Audard \etal\ 2007]{audard07}). The UV excess
supports an accretion scenario in which gas in accretion streams shock-heats near the surface
to form hot spots (\cite[Calvet \& Gullbring 1998]{calvet98}). The OM has also recorded a slow U-band
flux increase in a brown dwarf, most likely due to an ``accretion event'' covering a time span of
several hours (\cite[Grosso \etal\ 2007b]{grosso07b}).

Accretion shocks may leave signatures in line-dominated high-resolution X-ray spectra. Given the
typical mass accretion rates on T Tauri stars and their modest accretion hot spot filling factors
of no more than a few percent (\cite[Calvet \& Gullbring 1998]{calvet98}), densities of order $10^{12}$~cm$^{-3}$
or more are to be expected (\cite[Telleschi \etal\ 2007c]{telleschi07c}, \cite[G\"udel \etal\ 2007c]{guedel07c}),
and such densities are indeed indicated in the density-sensitive line ratios of O\,{\sc vii} and Ne\,{\sc ix}
triplets of some CTTS (e.g., \cite[Kastner \etal\ 2002]{kastner02}, \cite[Stelzer \& Schmitt 2004]{stelzer04}). XEST 
has added relevant information on two further accreting pre-main sequence stars: 
T Tau N (\cite[G\"udel \etal\ 2007c]{guedel07c}) and the Herbig star AB Aur 
(\cite[Telleschi \etal\ 2007b]{telleschi07c}). However, in both {\it XMM-Newton} RGS spectra,
the O\,{\sc vii} triplet line ratios are compatible with density upper limits of a few times $10^{10}$~cm$^{-3}$ 
(Fig.~\ref{fig:accretion}b), apparently not supporting the accretion-shock scenario. How important really is 
accretion for CTTS X-ray emission?

\subsection{The ``X-Ray Soft Excess''}

\begin{figure}
\begin{center}
 \includegraphics[angle=-0,width=11.cm]{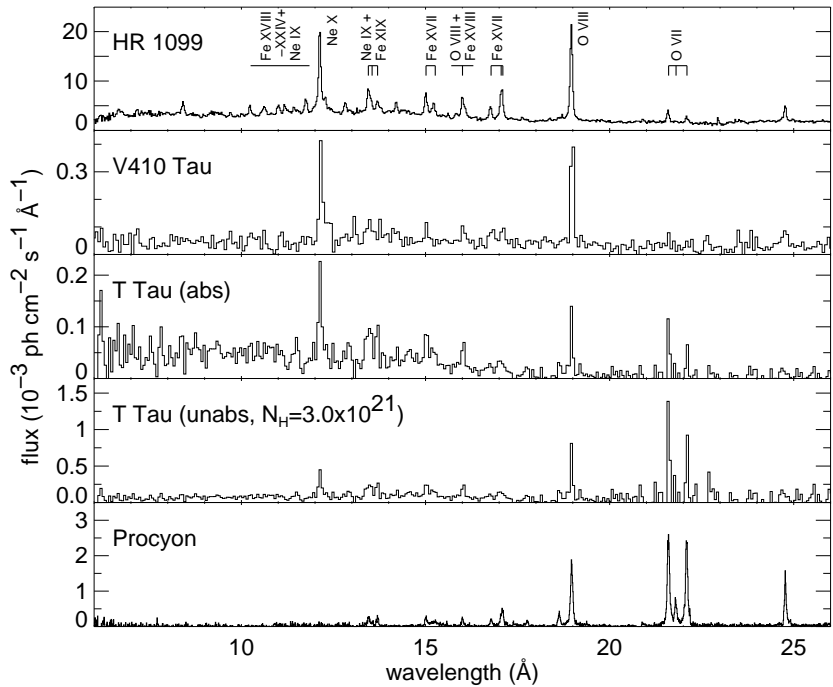}
 \caption{Comparison of fluxed {\it XMM-Newton} RGS photon spectra of (from top to bottom) the active binary HR~1099, 
         the WTTS V410~Tau, the CTTS T Tau, T Tau modeled after removal of absorption,  
	 and the inactive MS star Procyon. The bins are equidistant in wavelength.}\label{fig:ttau}
\end{center}
\end{figure}

Fig.~\ref{fig:ttau} compares {\it XMM-Newton} RGS spectra
of the active binary HR~1099 (X-rays mostly from a K-type subgiant; archival 
data) the weakly absorbed WTTS V410~Tau (\cite[Telleschi \etal\ 2007c]{telleschi07c}), 
the CTTS T Tau  (\cite[G\"udel \etal\ 2007c]{guedel07c}), and the old F subgiant Procyon (archival  
data). HR~1099 and V410~Tau show the typical signatures of a hot, active
corona such as a strong continuum, strong lines of Ne\,{\sc x} and of  highly-ionized Fe but little flux
in the O\,{\sc vii} line triplet. In contrast, lines of  C, N, and O dominate the soft spectrum of Procyon, 
the O\,{\sc vii} triplet exceeding the O\,{\sc viii} Ly$\alpha$ line in flux.  T Tau reveals
signatures of a very active corona shortward of 19~\AA\ but also an unusually strong
O\,{\sc vii} triplet. Because its $N_{\rm H}$ is large (in contrast to  $N_{\rm H}$ of
V410 Tau), we have modeled the intrinsic, unabsorbed
spectrum based on transmissions determined in XSPEC using $N_{\rm H}$
from EPIC spectral fits ($N_{\rm H} \approx 3\times 10^{21}$~cm$^{-1}$; \cite[G\"udel \etal\ 2007a]{guedel07a}). 
{\it The {\rm O\,{\sc vii}} lines are  the strongest lines in the intrinsic X-ray spectrum,} reminiscent of the 
situation in  Procyon!

\begin{figure}
\begin{center}
 \includegraphics[angle=-0,width=11cm]{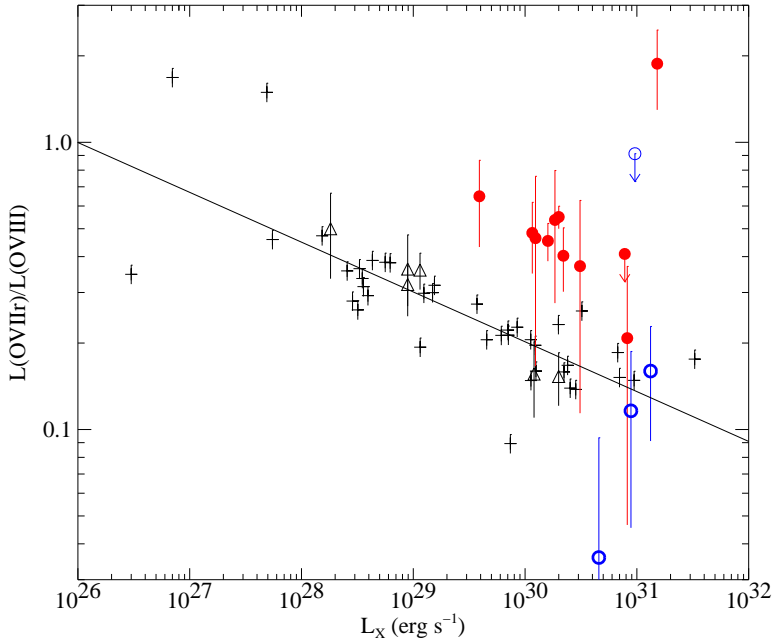}
 \caption{The ratio between O\,{\sc vii} $r$ and O\,{\sc viii} Ly$\alpha$ luminosities (each in erg~s$^{-1}$)
   vs. the total $L_{\rm X}$.  
   Crosses mark MS stars, triangles  solar analogs of
   different ages, filled (red) circles
   CTTS, and open (blue) circles WTTS. The solid line is a power-law fit to the MS stars (after
   G\"udel \& Telleschi 2007, submitted).}\label{fig:softexcess}
\end{center}
\end{figure}

To generalize this finding, we plot in Fig.~\ref{fig:softexcess} the ratio between the intrinsic (unabsorbed) 
luminosities of the O\,{\sc vii} $r$ line and the O\,{\sc viii} Ly$\alpha$ line as a function  of $L_{\rm X}$, 
comparing CTTS and WTTS with a larger MS sample (\cite[Ness \etal\ 2004]{ness04}) and MS solar analogs
(\cite[Telleschi \etal\ 2005]{telleschi05}). The TTS data 
are from \cite{robrade06}, \cite{guenther06}, \cite{argiroffi07}, \cite{telleschi07c}, and from 
our analysis of archival {\it XMM-Newton} data of RU Lup. For the TTS sample given 
by \cite{telleschi07c}, we have approximated $L$(O\,{\sc vii} $r$) $= 0.55L$(O\,{\sc vii}) 
(\cite[Porquet \etal\ 2001]{porquet01}). The trend for MS stars (black crosses and triangles, Sect.~\ref{accretors}) is 
evident: as the coronae get hotter toward higher $L_{\rm X}$, the ratio of O\,{\sc vii} $r$/O\,{\sc viii} 
Ly$\alpha$ line luminosities decreases. This trend is followed by the sample of WTTS,  while CTTS again 
show a  significant excess. This is the essence of the {\bf X-ray soft excess} in CTTS first discussed by 
\cite{telleschi07c} and \cite{guedel07c}: {\it Accreting pre-main sequence stars reveal a strong excess 
of cool (1-2~MK) material, regardless of the overall X-ray deficiency at higher temperatures.}

\subsection{Summary and conclusions on accretion-related X-ray emission}

Crucial XEST results on X-ray production and accretion can be summarized as follows:
\begin{enumerate}
\item[1.] Accreting CTTS show a general deficiency in X-ray production when referring
          to the hot coronal gas recorded by X-ray CCD cameras (\cite[Telleschi \etal\ 2007a]{telleschi07a});
\item[2.] the average coronal $T$ is higher in CTTS than in WTTS (\cite[Telleschi \etal\ 2007a]{telleschi07a});
\item[3.] All CTTS (except the two flaring sources SU Aur [also subject to high $N_{\rm H}$] and DH Tau) 
          show a {\bf soft excess} defined
          by an anomalously high ratio between the fluxes of the O\,{\sc vii} He-like triplet and the
          O\,{\sc viii}~Ly$\alpha$  line, when compared to WTTS and MS stars.
\end{enumerate}
The origin of the additional cool plasma in CTTS is likely to be related to the accretion
process.  Accretion streams may  shock-heat gas at the impact
point  to X-ray emitting temperatures. This model is supported by
high electron densities inferred from the observed O\,{\sc vii} or Ne\,{\sc ix} triplets in some
of the CTTS (e.g., \cite[Kastner \etal\ 2002]{kastner02}, \cite[Stelzer \& Schmitt 2004]{stelzer04}), 
although high densities were not seen in the two XEST accretors  T Tau (\cite[G\"udel \etal\ 2007c]{guedel07c}) 
and AB Aur (\cite[Telleschi \etal\ 2007b]{telleschi07b}). Alternatively, the cool, infalling material
may partly cool pre-existing heated coronal plasma, or reduce the efficiency of coronal heating in 
the regions of infall (\cite[Preibisch \etal\ 2005]{preibisch05}, \cite[Telleschi \etal\ 2007c]{telleschi07c},
\cite[G\"udel \etal\ 2007c]{guedel07c}). This model would at the same time  explain why CTTS are X-ray 
weaker than WTTS (\cite[Preibisch \etal\ 2005]{preibisch05}, \cite[Telleschi \etal\ 2007a]{telleschi07a}).  
We cannot assess what the relative importance of these processes is. It seems clear, however, that the 
soft excess described here argues in favor of a substantial influence of  accretion on the X-ray production
in pre-main sequence stars.

\section{Outflow-related X-rays}

The shock temperature in jets can be expressed as $T\approx 1.5\times 10^5v_{\rm 100}^2$~K 
(for fully ionized gas) where  $v_{100}$ is the shock  speed  in units of 
100~km~s$^{-1}$ (e.g., \cite[Raga \etal\ 2002]{raga02}). Jet speeds are typically of order $v = 300-500$~km~s$^{-1}$ 
(\cite[Eisl\"offel \& Mundt 1998]{eisloeffel98}, \cite[Anglada 1995]{anglada95}, 
\cite[Bally \etal\ 2003]{bally03}), in principle
allowing for shock speeds of similar magnitude. 

Faint, soft X-ray emission  has been detected from a few protostellar 
HH objects (\cite[Pravdo \etal\ 2001]{pravdo01}, \cite[Pravdo \etal\ 2004]{pravdo04}, 
\cite[Pravdo \& Tsuboi 2005]{pravdo05}, \cite[Favata \etal\ 2002]{favata02}, 
\cite[Bally \etal\ 2003]{bally03}, \cite[Tsujimoto \etal\ 2004]{tsujimoto04}, 
\cite[Grosso \etal\ 2006]{grosso06}).
\cite{bally03} used a {\it Chandra} observation to show that  X-rays 
form within an arcsecond of the protostar L1551 IRS-5 while the star itself
is too  heavily obscured to be detected. Strong absorption and extinction of
protostars and their immediate environment make the launching region of their powerful jets  
generally inaccessible to optical, near-infrared, or X-ray studies. However, a class of strongly 
accreting, optically revealed CTTS also exhibit so-called micro-jets 
visible in optical lines (\cite[Hirth \etal\ 1997]{hirth97}), with flow speeds similar to 
protostellar jets. CTTS micro-jets have  the 
unique advantage that they can - in principle - be followed down to the 
acceleration region both in the optical and in X-rays. 

\subsection{``Two-Absorber X-Ray'' (TAX) Sources}

\begin{figure}
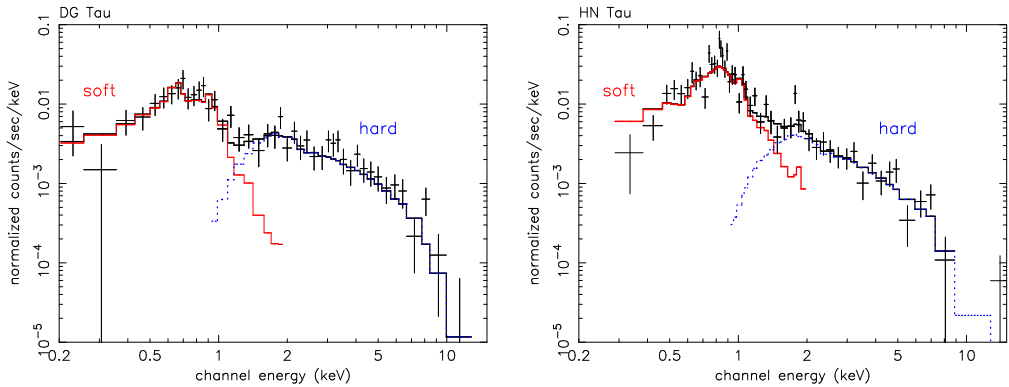

 {
 \includegraphics[angle=-90,width=6.4cm]{guedelfig4a.ps}
 \hskip 0.4truecm\includegraphics[angle=-90,width=6.4cm]{guedelfig4b.ps}
 }
 \vskip 0.4truecm
\caption{{\it XMM-Newton} EPIC PN spectra of the jet sources DG Tau (left) and HN Tau (right). The 
         red (solid) and blue (dotted) histograms show 
         the spectral fits pertaining to the soft and hard component, respectively (DG Tau after 
	 \cite[G\"udel \etal\ 2007b]{guedel07b}).}\label{fig:pinkspec}
\end{figure}

X-ray spectra of very strongly accreting, micro-jet driving CTTS exhibit an anomaly (Fig.~\ref{fig:pinkspec},
\cite[G\"udel \etal\ 2005]{guedel05}, \cite[G\"udel \etal\ 2007b]{guedel07b}):
The spectra of DG Tau, GV Tau, DP Tau, CW Tau, and HN Tau are composed of two components, a cool component
subject to very low absorption and a hot component subject to photoelectric absorption
about one order of magnitude higher. For similar phenomenology, see also \cite{kastner05}
and \cite{skinner06}.
The cool component shows temperatures atypical for T Tau stars, 
ranging from $\approx 3-6$~MK, while the hot component reveals extremely high temperatures
(10--100~MK).  We discuss in the following the best example, the single CTTS DG Tau.

The hard spectral component of the DG Tau point source is unusually strongly absorbed, with a gas column density
($N_{\rm H} \approx 2\times 10^{22}$~cm$^{-2}$) higher by a factor of $\approx 5$ than predicted from the visual 
extinction $A_{\rm V}$ of 1.5--3~mag if standard gas-to-dust ratios are assumed (see \cite[G\"udel \etal\ 2007b]{guedel07b} 
and references therein). Because the hard component occasionally flares, in one case being preceded by U band emission 
as in solar and stellar flares (\cite[G\"udel \etal\ 2007b]{guedel07b}), it is most straightforwardly 
interpreted as coronal or ``magnetospheric''.  The {\it excess absorption} is likely to be due to the heavy accretion streams
falling down along the magnetic fields and absorbing the X-rays from the underlying corona/magnetosphere. The 
{excess absorption-to-extinction} ($N_{\rm H}/A_{\rm V}$) ratio then is an indicator of dust sublimation:
{\it the accreting gas streams are dust-depleted}.

In contrast, $N_{\rm H}$  of the soft X-ray component,
$N_{\rm H} = 1.3~(0.7-2.4)\times 10^{21}$~cm$^{-2}$ (90\% error range)
is {\it lower} than suggested from the stellar $A_{\rm V}$, $N_{\rm H}(A_{\rm V}) 
\approx (3-6)\times 10^{21}$~cm$^{-2}$. A likely
origin of these X-rays is the base of the  jet. Such an origin is suggested by i) the unusually
soft emission not usually seen in T Tauri stars (\cite[G\"udel \etal\ 2007a]{guedel07a}), 
ii) the low $N_{\rm H}$, and
iii) the explicit evidence of jets in the {\it Chandra} image, as we will show below.

\subsection{X-rays and jets}

\begin{figure}
 \vskip -0.7truecm\hbox
 {
 \includegraphics[angle=-0,width=6.5cm]{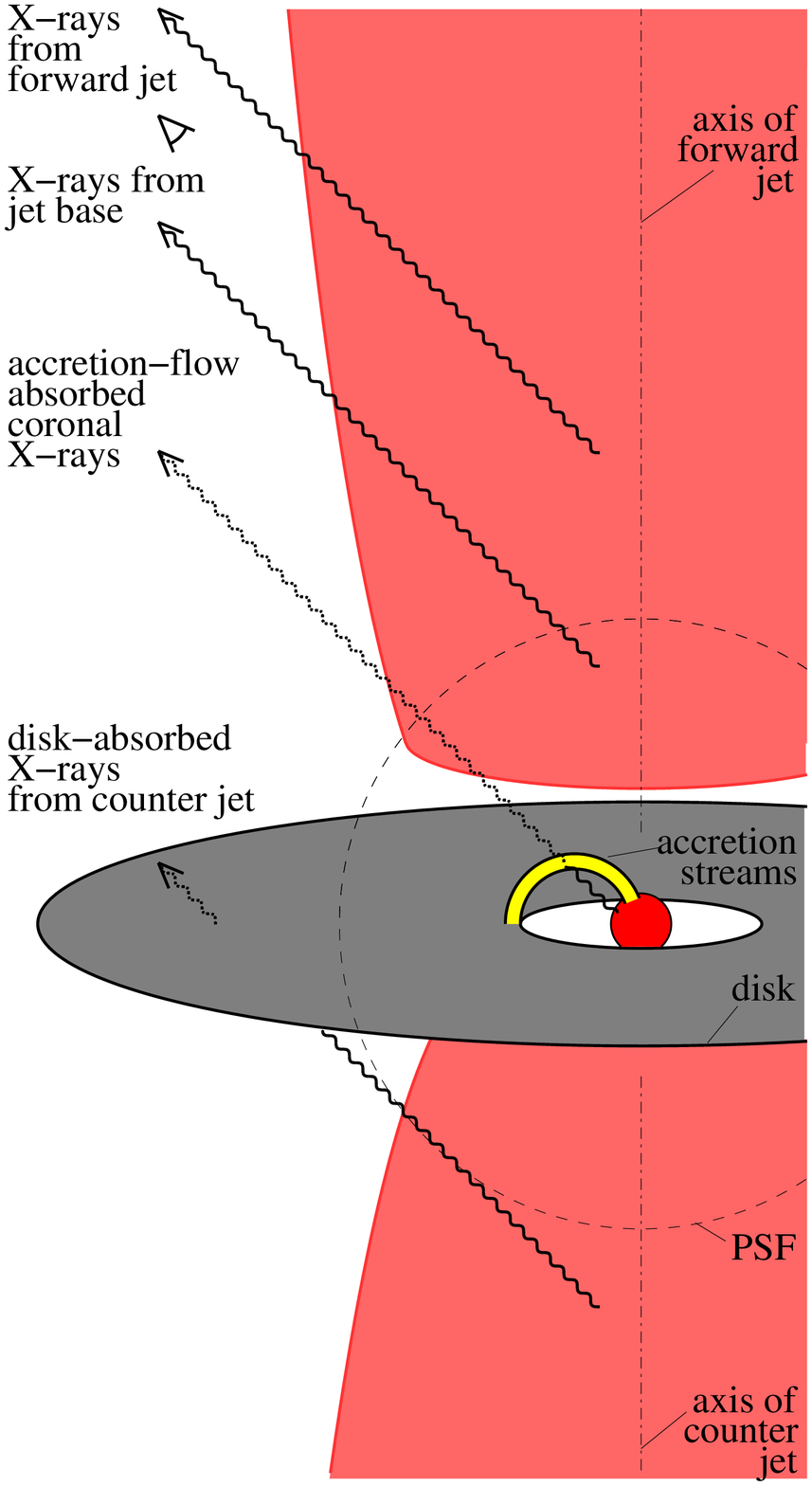} 
 \hskip 0.7truecm\vbox
 {
 \includegraphics[angle=-0,width=5.578cm]{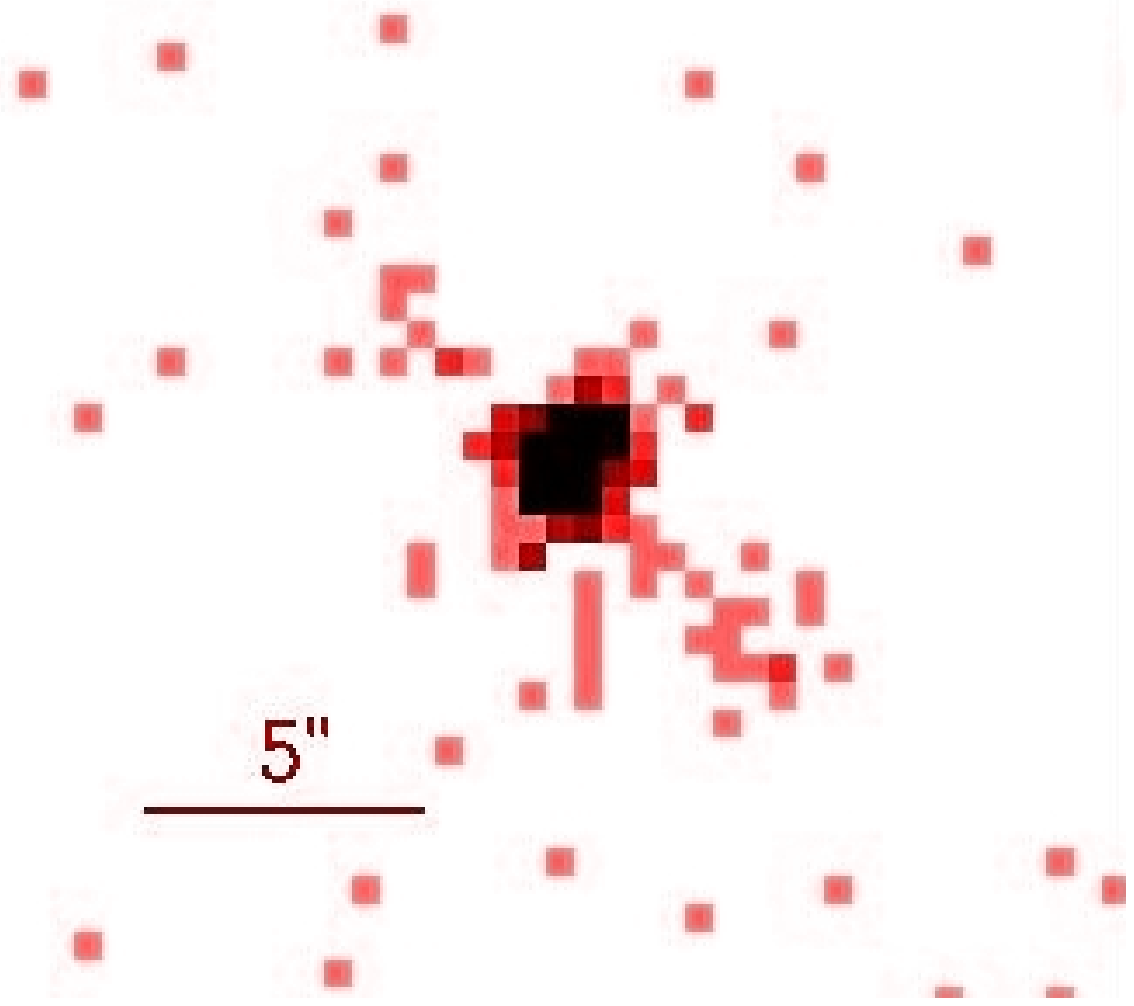} \\
 \includegraphics[angle=-0,width=5.578cm]{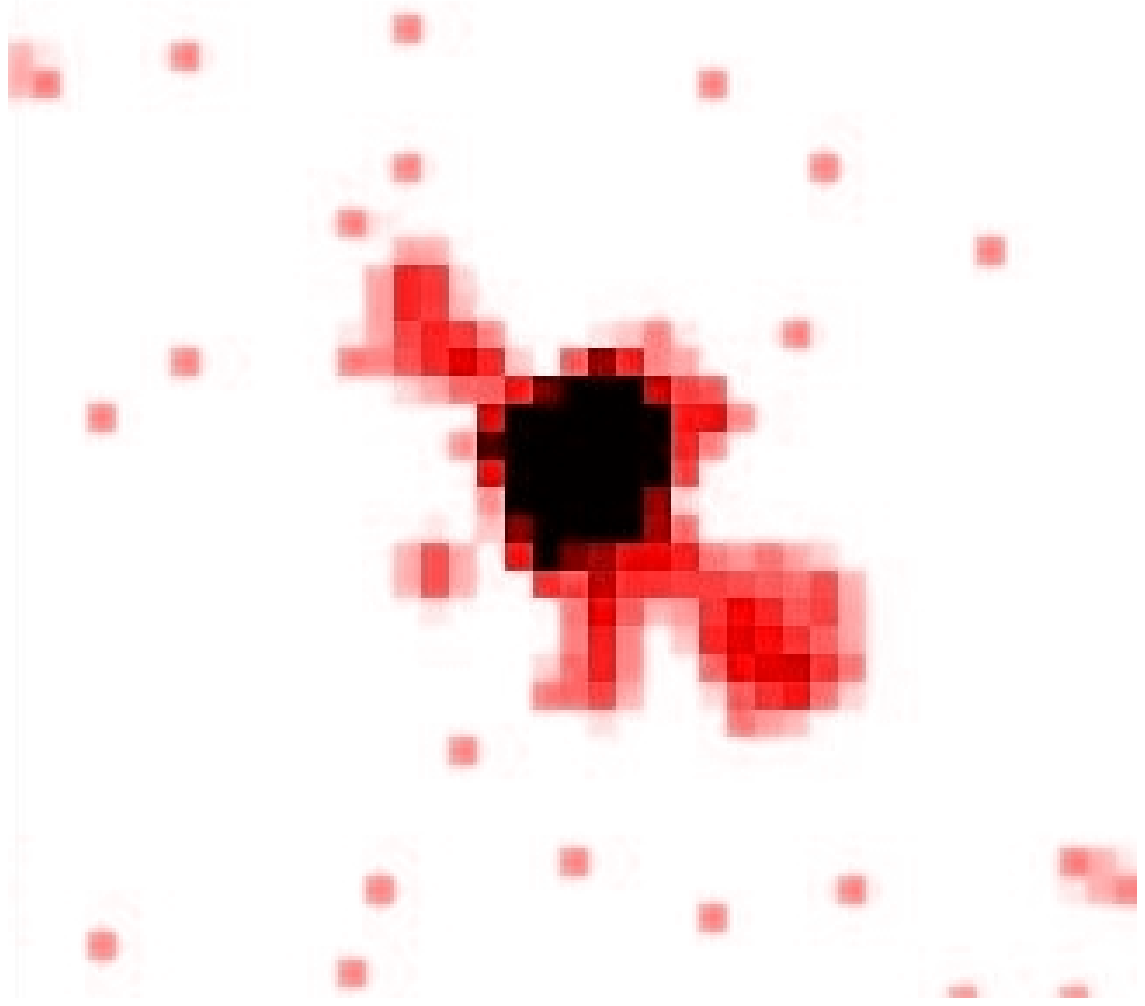} 
 }
 }
  \caption{{\it Left (a):} Proposed model for the disk-jet interface with various X-ray sources,
           including: the spatially resolved
	   forward jet; the spatially unresolved (within the stellar PSF) but spectrally resolved sources at the jet base; the accretion-stream
	   absorbed corona/magnetosphere; and the disk-absorbed counter-jet sources. -- {\it Right (b)}: Two {\it Chandra} ACIS-S
	   0.6--1.7~keV images  of DG Tau and its jets; the lower figure has been smoothed (after G\"udel \etal\ 2007d, submitted).}\label{fig:jetim}
\end{figure}

A {\it Chandra} X-ray image of the DG Tau environment is shown in Fig.~\ref{fig:jetim}b (pixel size
0.49$^{\prime\prime}$). This image
was produced by combining counts from a total of 90~ks of  {\it Chandra} exposure time. Also
shown is a smoothed version. To suppress background and to emphasize the soft sources, 
only counts within the 0.6--1.7~keV range are plotted. There is clear
evidence for a jet-like extension outside the stellar point spread function (PSF)
to the SW along a position angle of $\approx 225$~deg, but 
we also find  a significant excess of counts in the NE direction (PA $\approx 45$~deg). This is
coincident with the jet optical axis, which for the SW jet has been given as  217-237~deg 
(\cite[Eisl\"offel \& Mundt 1998]{eisloeffel98}). We verified, using raytrace
simulations, that the jet sources are extended: a faint point source would occupy only a few pixels.
We also find the counter jet to be harder, with photon energies mostly above 1~keV, while 
the forward jet shows a mixture of softer and harder counts. The spectral properties of the jet sources
are reminiscent of the {\it soft} component in the ``stellar'' spectrum.

\subsection{Summary on jets}

DG Tau is the prototype of a new class of jet-driving X-ray sources. It hosts at least four
X-ray sources of different origin and subject to different $N_{\rm H}$ (see
Fig.~\ref{fig:jetim}a), namely:
\begin{enumerate}
\item[1.] a weakly absorbed, diffuse, soft component along the forward-jet axis; 
\item[2.] an intermediately absorbed, diffuse, soft component along the counter-jet axis; 
\item[3.] a weakly absorbed, compact, non-variable, soft component (within the stellar PSF); 
\item[4.] a strongly absorbed, compact, flaring, hard component (within the stellar PSF). 
\end{enumerate}

The low $N_{\rm H}$ of the {\it forward jet} and of the {\it soft stellar} component 
suggests that the soft spectral emission originates from a region ``in front'' of the star.
We identify the soft component with X-ray emission from the base of the jets.
In contrast, the harder {\it counter jet} suggests
stronger absorption by the  extended gas disk.
A determination of the gas-to-dust ratio is in principle possible by measuring
the differential absorption and extinction of the two jets. 
Finally, the {\it hard stellar} component  is attributed to a (flaring) corona, with the
excess photoelectric absorption due to dust-depleted accretion gas streams.

The combined power of the resolved jets and the
unresolved soft spectral component is of order $10^{29}$~erg~s$^{-1}$, similar to 
the X-ray output of a moderate T Tauri star. This emission, distributed
above the accretion disk, may be  an important contributor to X-ray heating
and ionization of gaseous disk surfaces (\cite[Glassgold \etal\ 2004]{glassgold04}).
We speculate that protostellar jets in general develop the same kind of jet X-ray emission,
but these sources remain undetected close to the star because of strong photoelectric
absorption.

\begin{acknowledgments}
This work has been supported by the International Space Science Institute in Bern,
the Swiss National Science Foundation (AT, MA, MG: grants 20-66875.01, 20-109255/1, PP002--110504), 
ASI/INAF (Palermo group, grant ASI-INAF I/023/05/0), and NASA (MA, SS, DP: grants NNG05GF92G, GO6-7003).
This research is based on observations obtained with {\it XMM-Newton}, an ESA science mission
with instruments and contributions directly funded by ESA member states and the USA (NASA).  
The CXC X-ray Observatory Center is operated by the Smithsonian Astrophysical
Observatory for and on behalf of the NASA under contract NAS8-03060.
\end{acknowledgments}

\end{document}